\documentclass[conference]{IEEEtran}
\IEEEoverridecommandlockouts

\usepackage{cite}
\usepackage{amsmath,amssymb,amsfonts}
\usepackage{algorithm}
\usepackage{algorithmic}
\usepackage{graphicx}
\usepackage{textcomp}
\usepackage{xcolor}
\usepackage{booktabs}
\usepackage{float}
\usepackage{placeins}
\usepackage{dblfloatfix}
\usepackage{tikz}
\usetikzlibrary{arrows.meta, positioning, fit, calc}
\usepackage{hyperref}

\def\BibTeX{{\rm B\kern-.05em{\sc i\kern-.025em b}\kern-.08em
T\kern-.1667em\lower.7ex\hbox{E}\kern-.125emX}}

\begin{document}

\title{Hardware-Accelerated Line-Rate Bitstream Screening for Secure FPGA Reconfiguration  
\thanks{This work was supported by the McNair Junior Fellowship and the Office of Undergraduate Research at the University of South Carolina.}
}

\author{
\IEEEauthorblockN{Rye Stahle-Smith, 
Carter Antley, 
Jason D. Bakos, and 
Rasha Karakchi}
\IEEEauthorblockA{\textit{Dept. of Computer Science and Engineering},
\textit{University of South Carolina}, Columbia, SC, USA \\
\{ryes, ctantley\}@email.sc.edu, \{jbakos, karakchi\}@cec.sc.edu}
}

\maketitle

\begin{abstract}
As Field-Programmable Gate Arrays (FPGAs) scale in multi-tenant cloud and edge-AI environments, the configuration bitstream has become a critical, yet opaque, security boundary. Existing hardware Trojan detection methods often rely on trusted design artifacts or computationally intensive reverse-engineering, introducing prohibitive latencies in dynamic, "just-in-time" reconfiguration workflows. 

This paper presents BLADEI (\textit{Bitstream-Level Abnormality Detection for Embedded Inference}), a bitstream-level security framework designed for deployment-time screening of FPGA configurations without requiring source code, netlists, or vendor-specific tooling. BLADEI introduces a hybrid architecture that combines multi-scale byte-sequence learning with compact statistical representations to detect anomalous configurations directly from raw bitstreams. We implement the framework on a Xilinx PYNQ-Z1 system, demonstrating an end-to-end cloud-to-edge pipeline that enforces security prior to FPGA configuration.

Evaluating across 1,383 bitstreams, BLADEI achieves a macro F1-score of 0.91. However, our systems-level characterization reveals a "preprocessing wall": software-based feature extraction accounts for 92\% of the total 16.4-second latency, while model inference requires only 1.4 seconds. To address this bottleneck, we propose a streaming hardware-accelerated feature extraction engine designed for the FPGA programmable logic (PL). The evaluation shows that PL-based streaming engine can reduce feature-extraction latency to the millisecond range. This work positions bitstream-level screening as a first-class primitive and demonstrates that hardware-accelerated preprocessing is the key enabler for securing next-generation reconfigurable custom computing machines at line rate
\end{abstract}



\section{Introduction}
Field-Programmable Gate Arrays (FPGAs) are central to modern reconfigurable computing, spanning cloud acceleration and edge-AI inference \cite{maxfield2004design,wolf2004fpga,alfke2011fpga,karakchi2023napoly, karakchi2017dynamically, mal2016design,marchand2014low,chakraborty2013hardware}. In these environments, hardware functionality is defined by binary bitstreams that are frequently generated and deployed dynamically. However, this flexibility introduces a critical security boundary: malicious or tampered bitstreams can embed hardware Trojans that leak data, manipulate inference, or degrade reliability \cite{chakraborty2013hardware,benz2012bil, krieg2016malicious,surabhi2024feint,krieg2023reflections}.

The risks are amplified in multi-tenant and FPGA-as-a-Service (FaaS) deployments where bitstreams originate from untrusted sources \cite{kawser2025multitenant}. Existing security approaches are poorly aligned with these dynamic workflows. Design-time verification requires trusted RTL or netlists \cite{mal2016design}, while side-channel methods require golden reference devices and controlled environments \cite{kawser2025multitenant, elnaggar2022learning,elnaggar2018machine,ghimire2025goldenfree}. Reverse-engineering techniques \cite{benz2012bil,yoon2018bitstream, seo2018reverse, chakraborty2013hardware} and other bitstream-to-netlist tools \cite{krieg2016malicious,surabhi2024feint,krieg2023reflections}, incur high computational overhead and are often vendor-dependent, making them unsuitable for "just-in-time" screening prior to configuration.

To address these constraints, recent research has turned to machine learning to detect malicious circuits directly within bitstreams \cite{stahle2025realtime, boudjadar2025dynamic,hou2024hardware,zhou2025security}. Although sequence-based and NLP-inspired models have shown high detection accuracy \cite{boudjadar2025dynamic,pynq2024}, they are typically evaluated in servers-class offline environments. These studies often overlook the end-to-end latency and resource limitations of the embedded platforms that actually perform the reconfiguration.

\textbf{Our Discovery: The Preprocessing Wall.} In this work, we propose BLADEI, a deployment-time screening framework that operates on raw configuration data without vendor-specific tooling. By evaluating BLADEI on a Xilinx PYNQ-Z1 SoC, we identify a fundamental systems-level bottleneck: bitstream screening is not \textit{compute-bound} by model inference, but rather \textit{preprocessing-bound} by feature extraction. Our characterization reveals that software-based byte-level processing accounts for 92\% of total screening latency, creating a "preprocessing wall" that prevents near-line-rate security enforcement.

\textbf{Our Solution: DMA-Path Acceleration.} To overcome this bottleneck, we shift the security primitive from the CPU to the hardware datapath. We propose a Byte-Statistics Engine designed to operate in the FPGA programmable logic (PL) as a streaming module on the DMA path. This architecture computes byte-frequency histograms and statistical features in a single pass, overlapping screening with bitstream transfer. Preliminary results show that this hardware-accelerated approach significantly reduces  latency from seconds to milliseconds and positioning bitstream screening as a practical near zero-latency component of the FPGA reconfiguration workflow.

\section{BLADEI Detection Framework}
\begin{figure}[htbp]
\centering
\includegraphics[width=0.95\linewidth]{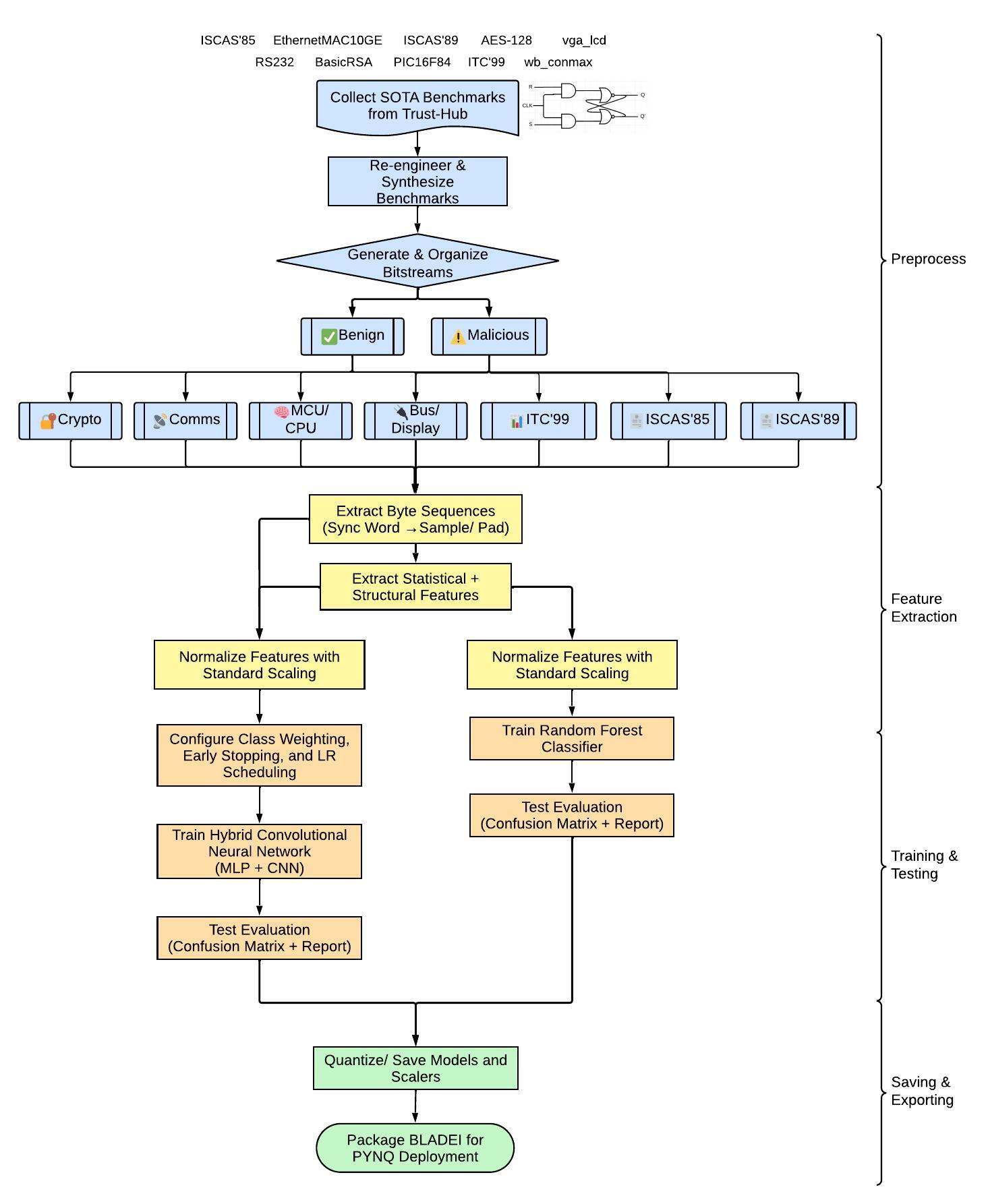}
\caption{Training and deployment workflow of BLADEI. The offline phase performs model training and validation, while the online phase executes on the target platform for deployment-time screening.}
\label{fig:fig3}
\end{figure}

BLADEI is designed as a deployment-time screening pipeline that operates on raw configuration data. The system is divided into an \textit{offline phase} for model training on a host system shown in Figure \ref{fig:fig3} and an \textit{online phase} for lightweight inference on the target platform.

\subsection{Bitstream Representation}
\begin{figure}[htbp]
\centering
\includegraphics[width=0.85\linewidth]{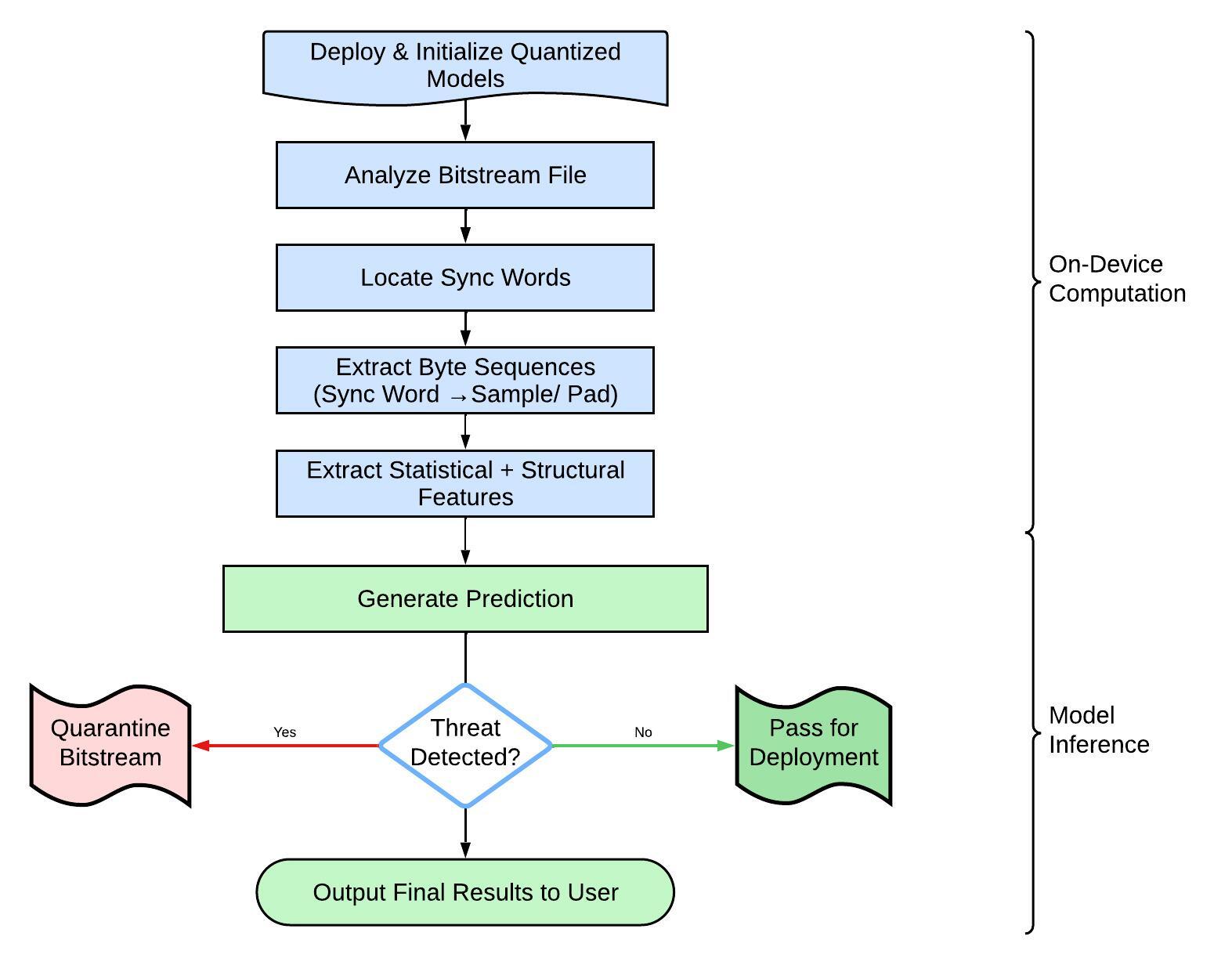}
\caption{Runtime screening pipeline on the PYNQ-Z1 platform. Incoming bitstreams are processed locally prior to FPGA configuration, enabling autonomous deployment-time security enforcement.}
\label{fig:fig4}
\end{figure}

Each FPGA bitstream is modeled as a raw byte sequence $B = [b_1, b_2, \dots, b_L]$, where $bi \in \{0, \dots, 255\}$. The configuration payload is isolated by identifying the vendor-specific synchronization word. To ensure consistent input for the neural network, a fixed-length segment of 4,096 bytes is sampled or padded. This "black-box" representation avoids reliance on proprietary vendor bitstream specifications. Figure \ref{fig:fig4} shows the runtime screening on the PYNQ-Z1 platform. 

\subsection{Dual-Branch Feature Extraction}
BLADEI utilizes a hybrid feature set to capture both local structural signatures and global statistical properties:

\begin{enumerate}
    \item \textbf{Sequence Features:} The byte sequence is processed by a 1D-Convolutional Neural Network (CNN). Multi-scale kernels capture localized patterns associated with placement and routing signatures unique to Trojan insertions.
    \item \textbf{Statistical Features:} Global bitstream characteristics are summarized in a 278-dimensional vector. This includes a normalized byte-frequency histogram, where each bin $v_j$ is defined as:
    \begin{equation}
        v_j = \frac{1}{L} \sum_{i=1}^{L} \mathbf{1}\{b_i = j\}, \quad j \in \{0, \dots, 255\}
    \end{equation}
\end{enumerate}

The vector is augmented with higher-order statistics, including entropy ($H$), variance ($\sigma^2$), and byte transition rates to capture structural density.

\subsection{Hybrid Detection Model}
The features are fed into a dual-head architecture. The CNN branch produces a 512-dimensional embedding that is concatenated with the statistical vector and passed to a binary classifier for Trojan detection. Similarly, a Random Forest head utilizes the statistical features to perform \textit{family classification}, identifying the design provenance (e.g., CRYPTO, MCU, BUS). This dual-task approach provides both security enforcement and workload characterization in a single pass.

Although the hybrid model provides high detection accuracy, its reliance on dense byte-level statistical features introduces a significant computational burden when executed in software. To bridge the gap between high-level detection logic and the strict latency requirements of real-time FPGA reconfiguration, we propose a specialized hardware architecture. The following section details the design of the \textit{Byte-Statistics Engine}, which offloads these intensive computations to the programmable logic to achieve line-rate performance.

To overcome the ``preprocessing wall'' identified in our systems characterization, we shift the feature extraction primitive from the CPU to a streaming hardware datapath. This section details the architecture of the proposed Byte-Statistics Engine.

\subsection{Architectural Overview}
\begin{figure}[htbp]
\centering
\includegraphics[width=0.95\linewidth]{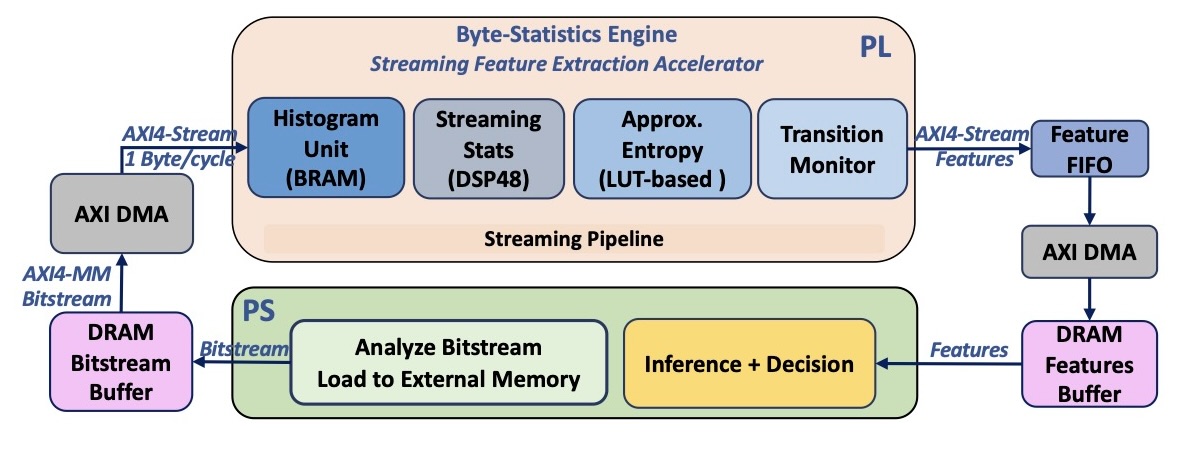}
\caption{The proposed system integrates a streaming Byte-Statistics Engine within FPGA programmable logic (PL). Bitstreams are transferred from DRAM via AXI DMA and processed at one byte per cycle using a fully pipelined architecture. Extracted features are written back to DRAM and consumed by the ARM processor for classification, enabling near line-rate preprocessing.}
\label{fig:pl}
\end{figure}

Figure \ref{fig:pl} shows the engine that is designed to operate within the FPGA Programmable Logic (PL) as a non-intrusive observer on the DMA data path. Using an AXI4-Stream interface, the engine snoops bitstream data at line rate (1 byte/cycle) as it is transferred from memory to the configuration controller. This allows statistical feature calculation to overlap entirely with the bitstream loading phase.

\section{Hardware-Accelerated Engine Design}
\label{sec:hardware_design}

\subsection{Functional Modules}
The engine consists of four primary modules that operate in parallel to ensure near-zero overhead through fully pipelined streaming:

\begin{itemize}
    \item \textbf{Histogram Unit:} This module utilizes a Dual-Port Block RAM (BRAM) configured as 256 bits to store byte frequencies ($v_j$). For each incoming byte $b_i$, the unit performs a read-increment-write operation. To maintain strict 1-byte-per-cycle throughput, the unit employs internal bypass logic to handle read-after-write hazards that occur from consecutive identical bytes.
    \item \textbf{Streaming Statistics:} To compute the mean and variance without a second pass over the data, this block utilizes DSP48 slices to maintain running totals of the byte values ($\sum b_i$) and their squares ($\sum b_i^2$).
    \item \textbf{Approximate Entropy:} The Shannon entropy ($H = -\sum p_i \log_2 p_i$) is approximated using a hardware-efficient approach. Since real-time logarithm calculation is resource-intensive, the engine utilizes a pre-computed Look-Up Table (LUT) to store $x \log_2 x$ values, which are accumulated as the bitstream stream concludes.
    \item \textbf{Transition Monitor:} A register-based comparator tracks the transition rate by detecting changes between consecutive bytes ($b_i \neq b_{i-1}$). This provides a metric for the structural density and complexity of the configuration file.
\end{itemize}

Once the final byte of the bitstream is processed, the resulting 278-dimensional feature vector is latched into a FIFO buffer. The extracted feature vector is streamed via an AXI-Stream interface to a DMA engine, which transfers the data into DRAM. The ARM processor reads the feature vector from memory for classification.

\section{Dataset and Evaluation}
\label{sec:evaluation}

We evaluate the \textit{BLADEI} framework from a system-level perspective, focusing on its ability to detect malicious configurations under realistic deployment constraints on embedded FPGA platforms.

\subsection{Dataset Construction and Augmentation}
The lack of large-scale, publicly available bitstream datasets necessitates a robust synthetic generation pipeline. We construct our evaluation corpus using 23 benchmarks from the \textbf{Trust-Hub} suite \cite{krieg2016malicious}, re-engineered for the Xilinx PYNQ-Z1 platform using Vivado v2023.2. 

The dataset spans seven distinct design families: \textit{CRYPTO, COMMS, MCU/CPU, BUS/DISPLAY, ITC99, ISCAS89, and ISCAS85}. To ensure that the model learns intrinsic Trojan signatures rather than implementation artifacts, we employ \textbf{diversified implementation variants}. By applying varying place-and-route directives, we generate structurally distinct bitstreams for the same functional design. The final augmented dataset comprises \textbf{1,383 bitstreams}, containing both benign and Trojan-inserted variants.

\begin{table}[h]
\centering
\caption{Cross-Validation Performance: Model Comparison}
\label{tab:comparison}
\begin{tabular}{lcccc}
\hline
\textbf{Model} & \textbf{Acc.} & \textbf{Prec.} & \textbf{Rec.} & \textbf{F1} \\ \hline
\textbf{CNN (hybrid)} & \textbf{0.91 $\pm$ 0.02} & \textbf{0.91 $\pm$ 0.02} & \textbf{0.91 $\pm$ 0.02} & \textbf{0.91 $\pm$ 0.02} \\
RF & 0.90 $\pm$ 0.02 & 0.90 $\pm$ 0.02 & 0.90 $\pm$ 0.02 & 0.90 $\pm$ 0.02 \\
GB & 0.90 $\pm$ 0.02 & 0.90 $\pm$ 0.03 & 0.90 $\pm$ 0.02 & 0.90 $\pm$ 0.02 \\
AB & 0.89 $\pm$ 0.03 & 0.90 $\pm$ 0.03 & 0.89 $\pm$ 0.02 & 0.89 $\pm$ 0.03 \\
LR & 0.85 $\pm$ 0.03 & 0.87 $\pm$ 0.02 & 0.83 $\pm$ 0.03 & 0.84 $\pm$ 0.03 \\
NB & 0.61 $\pm$ 0.04 & 0.58 $\pm$ 0.04 & 0.58 $\pm$ 0.04 & 0.58 $\pm$ 0.04 \\
SVM (RBF) & 0.68 $\pm$ 0.02 & 0.69 $\pm$ 0.07 & 0.62 $\pm$ 0.02 & 0.62 $\pm$ 0.03 \\
KNN & 0.90 $\pm$ 0.03 & 0.89 $\pm$ 0.03 & 0.90 $\pm$ 0.03 & 0.89 $\pm$ 0.03 \\
DT & 0.91 $\pm$ 0.03 & 0.90 $\pm$ 0.04 & 0.90 $\pm$ 0.03 & 0.90 $\pm$ 0.04 \\ \hline
\end{tabular}
\end{table}

\subsection{Detection Performance}
We evaluated the effectiveness of the hybrid CNN-based architecture using a categorized 80/20 train-test split. As shown in Table \ref{tab:comparison}, the \textit{BLADEI} hybrid model achieves a \textbf{macro F1-score of 0.91}, outperforming standard machine learning baselines such as SVM and Naive Bayes. The confusion matrix reveals high sensitivity, with 127 true positives and only 7 false negatives. Additionally, the Random Forest head achieves near-perfect accuracy in design family classification, confirming that the extracted statistical features capture meaningful structural provenance.

\subsection{Deployment-Time Latency Analysis}
A core contribution of this work is the characterization of the \textbf{``Preprocessing Wall.''} We measured the end-to-end latency on the PYNQ-Z1 processing system, decomposing the workflow into loading, feature extraction, and inference.

\begin{itemize}
    \item \textbf{Total Latency:} Average 16.4 seconds per bitstream.
    \item \textbf{Inference Time:} Model prediction requires only 1.3--1.4 seconds.
    \item \textbf{The Bottleneck:} Software-based feature extraction accounts for \textbf{92\% of total latency} ($\sim$15.0s).
\end{itemize}

Table \ref{tab:timing} reports the classification versus the timing of the trial in PYNQ-Z1. The result deployment-time screening is not compute-bound by the neural network, but rather throughput-bound by byte-level data processing.

\begin{table}[t]
\centering
\caption{Classification and Timing per Trial on PYNQ-Z1}
\label{tab:timing}
\resizebox{\columnwidth}{!}{
\begin{tabular}{clllcccc}
\hline
\textbf{Trial} & \textbf{Act. Cls.} & \textbf{Pred. Cls.} & \textbf{Fam.} & \textbf{Bench.} & \textbf{Load (ms)} & \textbf{Extr. (s)} & \textbf{Pred. (s)} \\
\hline
1 & Mal. & Mal. & CRYPTO & AES & 16.0 & 14.99 & 1.37 \\
2 & Ben. & Ben. & ISCAS89 & s15850 & 16.2 & 15.01 & 1.37 \\
3 & Ben. & Ben. & CRYPTO & BasicRSA & 16.48 & 14.95 & 1.57 \\
4 & Mal. & Mal. & ITC99 & b19 & 16.32 & 14.91 & 1.36 \\
5 & Mal. & Mal. & ISCAS85 & c1908 & 16.42 & 14.96 & 1.37 \\
\hline
\end{tabular}
}
\end{table}

\subsection{Hardware Utilization}
BLADEI Implementation runs on the processing system (ARM) of the Xilinx PYNQ-Z1 platform \cite{pynq2024, vohra2019pynqtorch}. The resources of the FPGA fabric utilized during the operation reflect the overlay of the base system and the overhead of the PYNQ environment. The hardware resource values presented were obtained using Vivado by synthesizing and implementing the base project files corresponding to PYNQ OS v2.4. These metrics are detailed in Table~\ref{tab:resources}.

\begin{table}[h]
\centering
\caption{Hardware Resource Utilization on PYNQ-Z1 (Running PYNQ OS v2.4)}
\label{tab:resources}
\begin{tabular}{lccc}
\hline
\textbf{Resource} & \textbf{Used} & \textbf{Available} & \textbf{Util.} \\ \hline
Slice LUTs & 33,109 & 53,200 & 62.23\% \\
Slice Registers & 47,370 & 106,400 & 44.52\% \\
Block RAM & 63 & 140 & 45.00\% \\
DSPs & 18 & 220 & 8.18\% \\
Bonded IOB & 118 & 125 & 94.40\% \\
Slices & 12,886 & 13,300 & 96.89\% \\ 
 On-chip power & \multicolumn{3}{c}{2.343 W}\\ \hline
\end{tabular}
\end{table}










\begin{table}[t]
\centering
\caption{Software vs. Ideal vs. HLS Kernel Performance}
\label{tab:perf_compare}
\begin{tabular}{lccc}
\hline
\textbf{Metric} & \textbf{Software (CPU)} & \textbf{Ideal PL} & \textbf{HLS Kernel} \\
\hline
Throughput (MB/s) & 0.26 & 100 & $\sim$84 \\
Latency (3.86 MB) & 15.0 s & 38.6 ms & $\sim$46 ms \\
Speedup & 1$\times$ & $\sim$385$\times$ & $\sim$320$\times$ \\
\hline
\end{tabular}
\end{table}

\subsection{Hardware Acceleration}
To address the identified preprocessing bottleneck, we evaluate the performance of a streaming hardware implementation. The software-based pipeline processes an average bitstream ($S_{avg}$) of 3.86~MB in approximately 15.0~seconds. The software throughput ($T_{sw}$) is defined as:

\begin{equation}
T_{sw} = \frac{S_{avg}}{L_{sw}} = \frac{3.86\text{ MB}}{15.0\text{ s}} \approx 0.26\text{ MB/s}
\end{equation}

We propose a Byte-Statistics Engine implemented in FPGA Programmable Logic (PL) operating at a clock frequency ($f$) of 100~MHz. Designed to process $N=1$ byte per clock cycle, the ideal peak streaming rate ($T_{hw}$) is:

\begin{equation}
T_{hw} = f \times N = 100\text{ MHz} \times 1\text{ byte/cycle} = 100\text{ MB/s}
\end{equation}

The ideal speedup ($S$) is calculated as the ratio of hardware to software throughput:

\begin{equation}
S = \frac{T_{hw}}{T_{sw}} = \frac{100\text{ MB/s}}{0.26\text{ MB/s}} \approx 385\times
\end{equation}

Accounting for potential DMA overhead and bus contention, we conservatively project a $300\times$ speedup. Under this model, the feature extraction latency is reduced to:

\begin{equation}
L_{hw} = \frac{S_{avg}}{T_{hw}} = \frac{3.86\text{ MB}}{100\text{ MB/s}} = 38.6\text{ ms}
\end{equation}

Table~\ref{tab:perf_compare} summarizes the comparison between the measured software implementation, the ideal streaming datapath, and the HLS-based kernel estimate. While the ideal architecture achieves a peak throughput of 100~MB/s under a fully pipelined 1-byte-per-cycle model, the HLS kernel achieves an effective throughput of approximately 84~MB/s due to fixed overheads including histogram initialization, reduction, and feature emission.

Despite this overhead, the HLS results closely track the ideal streaming model and validate the proposed hardware architecture. The reduction in feature extraction latency from seconds to tens of milliseconds demonstrates that preprocessing can be effectively overlapped with bitstream transfer, enabling near-line-rate deployment-time screening.

\section{Conclusion and Future Work}
\label{sec:conclusion}

This paper presented \textit{BLADEI}, a deployment-time security framework for detecting malicious hardware Trojans directly from raw FPGA bitstreams (BLADEI is publicly available in \cite{bread2025pynq}). By implementing the framework on a Xilinx PYNQ-Z1 system, we provided a comprehensive systems-level characterization of bitstream screening in edge environments. Our analysis identified a critical ``Preprocessing Wall,'' where software-based feature extraction accounts for 92\% of total screening latency, creating a significant bottleneck for dynamic reconfiguration workflows. 

To overcome this, we proposed a streaming \textbf{Byte-Statistics Engine} designed to operate on the DMA data path. The results demonstrate that this hardware-accelerated approach can achieve significant speedup, reducing preprocessing latency from seconds to approximately milliseconds. This transformation enables bitstream screening to function as a near-zero-overhead primitive, securing the cloud-to-edge pipeline at line rate.

Future research will focus on the following directions:
\begin{itemize}
    \item \textbf{System Implementation and Validation:} We plan to implement and synthesize the complete system to measure actual power consumption and area overhead on diverse FPGA architectures.
    \item \textbf{Cross-Vendor Generalization:} While this work focused on Xilinx bitstreams, we aim to extend the hybrid CNN architecture to support Intel and Lattice configurations, evaluating the model's ability to learn vendor-agnostic structural signatures.
    \item \textbf{Adversarial Robustness:} We intend to investigate the resilience of the \textit{BLADEI} framework against adversarial bitstream obfuscation techniques, such as bit-flipping attacks designed to manipulate statistical entropy without altering functional logic.
\end{itemize}



\bibliographystyle{IEEEtran}
\bibliography{main}

\end{document}